\documentclass[conference]{IEEEtran}
\IEEEoverridecommandlockouts

\usepackage[
backend=biber,
style=numeric,
sorting=ynt,
]{biblatex}
\usepackage{amsmath,amssymb,amsfonts}
\usepackage[english]{babel}
\usepackage[utf8]{inputenc}
\usepackage{algorithm}
\usepackage[noend]{algpseudocode}
\usepackage{graphicx}
\usepackage{textcomp}
\usepackage{csquotes}
\usepackage{xcolor}
\usepackage{adjustbox}
\usepackage{subfigure}
\def\BibTeX{{\rm B\kern-.05em{\sc i\kern-.025em b}\kern-.08em
    T\kern-.1667em\lower.7ex\hbox{E}\kern-.125emX}}
\begin{document}

\title{Effect of Miner Incentive on the Confirmation Time of Bitcoin Transactions  

}

\author{\IEEEauthorblockN{Befekadu G. Gebraselase, Bjarne E. Helvik, Yuming Jiang}
\IEEEauthorblockA{\textit{Department of Information Security and Communication Technology} \\
\textit{NTNU, Norwegian University of Science and Technology, Trondheim, Norway}\\
\{befekadu.gebraselase, bjarne, yuming.jiang\}@ntnu.no}

}

\maketitle

\begin{abstract}
Blockchain is a technology that provides a distributed ledger that stores previous records while maintaining consistency and security. Bitcoin is the first and largest decentralized electronic cryptographic system that uses blockchain technology. It faces a challenge in making all the nodes synchronize and have the same overall view with the cost of scalability and performance. In addition, with miners' financial interest playing a significant role in choosing transactions from the backlog, small fee or small fee per byte value transactions will exhibit more delays.   To study the issues related to the system's performance, we developed an $M(t)/M^N/1$ model.  The backlog's arrival follows an inhomogeneous Poison process to the system that has infinite buffer capacity, and the service time is distributed exponentially, which removes $N$ transactions at time. Besides validating the model with measurement data, we have used the model to study the reward distribution when miners take transaction selection strategies like fee per byte, fee-based, and FIFO. The analysis shows that smaller fee transactions exhibit higher waiting times, even with increasing the block size. Moreover, the miner transaction selection strategy impacts the final gain. 
\end{abstract}


\section{Introduction}
Cryptocurrency, which is a digital equivalence of fiat currency, is becoming popular. As of April 2020, there were approximately 5,392 cryptocurrencies being traded with a total market capitalisation of 201 billion dollars\footnote{https://finance.yahoo.com/news/top-10-cryptocurrencies-market-capitalisation-160046487.html}. The volume of transactions and circulation of these cryptocurrencies are uneven \cite{crypto1}. As of April 2020, Bitcoin (BTC), ether (ETH) and Ripple (XRP) were the top three cryptocurrencies with market capitalization of 128, 19.4, and 8.22 billion dollars respectively. Bitcoin is an autonomous decentralized virtual currency that removes the intermediary between participating parties while the cryptographic encryption and peer-to-peer formations provide the security.  This property has attracted much attention from the research and industry world to develop and integrate blockchain in the supply-demand chain.  The amount of Bitcoin usage and integration exhibits rapid increases in recent years.  For instance, the number of transactions per day in 2020 is twice higher as from 2016 to 2018 \cite{Btc}. 


Bitcoin has become popular with an increasing number of transaction requests over time. However, due to the limited capacity by design (average one block per 10 minutes) of Bitcoin, the number of transactions that the system can handle is also limited. This necessitates a strategy for a miner to select transactions in forming blocks. While it is natural for the miners to priority higher fee transactions to gain financially, such a strategy may cause a long delay in transaction confirmation for lower fee transactions. As a consequence, such financial gain oriented strategies may reduce the overall quality of the services provided by the ledger.  Furthermore, as the number of users increases unexpectedly while the number of mining nodes and pools rises linearly \cite{mingpool}, this makes Bitcoin unsuitable for small fee transactions.

Bitcoin is facing criticism over the scalability and performance \cite{challenges}.  It is imperative to study Bitcoin's transaction confirmation process' characteristics since they are critical indicators of how scalable the ledger is \cite{feeimportance}. To this end, some models have been proposed to study the average waiting time seen by transactions while considering the coefficient of arrival variation, batch processing, and block sizes \cite{BlockchainQueueingTheory}\cite{DiscreteBlockchain}.  However, based on a recent measurement-based work reported in \cite{TransBitcoin}, it is found that the transaction arrivals  follow an  inhomogeneous Poisson process and the arrival attributes have week correlations. In addition, the fee per byte is the default ordering mechanism in Bitcoin, while not just fee \cite{TransactionConfirmationBitcoin}. However, this fact is not addressed by most of the available modeling works, including  \cite{trasactionConfirmation}\cite{modlingBlockchain}.  In this paper, we consider these insights to model and study the transaction waiting time. 

This paper aims to investigate how different transaction selection strategies may affect the performance of Bitcoin in terms of transaction waiting time. However, there is a challenge: We cannot widely introduce such a strategy on Bitcoin. (i) For this reason, we develop a queueing model that simulates the behavior of Bitcoin with a focus on transaction waiting time. In the literature, several queueing models have been proposed. Our work proposes a new queueing model based on our previous extensive investigation on transaction handling and characteristics of Bitcoin. Based on this queueing model, a simulator is developed. The model/simulator is validated with measurement data from Bitcoin. (ii) With the simulator, we then study the transaction waiting time under different transaction selection/scheduling strategies, which include (Bitcoin default) fee per byte and fee-based. 
Beside this, we also consider the impact of increasing the block size on transaction waiting time. (iii) In addition, to account for that different miners may adopt different strategies, an investigation is also provided to check potential gain or loss to a miner.


The rest of the paper is organized as follows. The current state of the art is covered in Section \ref{sec-stateArt}. 
Following that, Section \ref{sec-model} presents the queueing model description and the simulator workflow, and validates the model results. After that, experimental results are discussed in Section \ref{sec-res}. Next, Section \ref{sec-rew} presents results from comparing different strategies. Section \ref{sec-dis} opens up a discussion on what has been observed in the analysis. Finally, Section \ref{sec-con} concludes the paper and outlines future research extensions.      


\section{Related Work} \label{sec-stateArt}

\subsection{Queueing Models of Transaction Waiting / Confirmation Time}
There are several works related to studying the average waiting time of transactions before their confirmations using queueing models. S.  Geissler et al. \cite{DiscreteBlockchain} proposed a $GI/GI^N/1$ model where the inter-arrival and batch service times follow an independent general distribution. Based on the model, they were able to show that the average arrival intensity variations and block size play a significant role in the confirmation times. Similarly, Lie et al. \cite{BlockchainQueueingTheory} illustrated that the block size and average arrival intensity exhibiting a significant factor in the average waiting time by developing a $GI/M^N/1$, where the inter-arrival time follows a general distribution but the batch service time follows an exponential distribution. 

Yoshiaki Kawase and Shoji Kasahara \cite{TransactionConfirmationBitcoin} developed an $M/G^B/1$ model where a batch service is used to reflect the block size limitations with the arrivals being blocked from entering into a block under the mining phase. The sojourn time of a transaction corresponds to its confirmation time. The same authors \cite{trasactionConfirmation} showed that because of a high arrival intensity, even high fee transactions are exhibiting a higher average waiting time.  Additionally, it was observed how the low fee and block size significantly affect transaction confirmation time. Similarly, ~\cite{batchqueuing} developed a batch processing queueing system that uses numerical and trace-driven simulation to validate exponential distribution, and hyper-exponential one can accurately estimate the mean transaction-confirmation time for the legacy 1MB block size limit. 

Mišić et al. \cite{networkDelivery} developed an analytical model to capture the Bitcoin P2P network. They developed a priority-based queuing model ($M/G/1$) of Bitcoin nodes and a Jackson network model of the whole network.  The study illustrated that the block size, node connectivity, and the overlay network significantly affect the probability of fork occurrence.  Furthermore, the study demonstrated the data distribution in the P2P network is sub-exponential, and the transaction traffic has less effect on the block propagation traffic mainly because of the priority. Motlagh et al. \cite{churn} developed a Continuous Time Markov Chain  (CTMC) model to study the churning process of a node with a homogeneous sleep time.  The analysis shows that results indicate that sleep times of the order of several hours require synchronization times in the order of a minute. 

Most of the research mentioned above works to evaluate the blockchain technology's performance concerning block size, transactions, node connectivity, churn, and block delivery. However, little has been investigated about the impact of the transaction selection strategy in forming blocks, considering the weak dependency between transaction attributes, and the inhomogeneous transaction arrivals.

\subsection{Reward Distribution}
Salimitari et al. \cite{profitMin} developed a prospect theoretical model to predict what a miner can mine relative to its hash rate power and electricity costs and how much may be expected to make from each pool. It was also demonstrated that the best pool for a miner to join is not always the same for all.  Liu et al. \cite{miningstrategy} proposed to introduce a forwarding node to reduce time delay for message propagation and increase the probability for a new block to be appended on the longest blockchain.  Samiran et al. \cite{bwh} performed an analysis on how a selfish miner could earn some extra incentive for launching a block withholding attack on a mining pool. This additional incentive comes from some other like-minded mining pool that wants to benefit from this block withholding attack. A. Laszka, B. Johnson, and J. Grossklags \cite{mindry} developed a game-theoretical model to study the impact of attacks on mining pools in either short or long-term effects.  This model is used to consider when the miner has an incentive to attack the pool or has no incentives to conduct the attack. Eyal \cite{minerDilama} showed that identical mining pools attack each other. They have demonstrated no Nash equilibrium when there is no attack on the pool; this will increase earned by participating parties. When two pools can attack each other, they face a version of the Prisoner’s Dilemma. If one pool chooses to attack, the victim’s revenue is reduced, and it can retaliate by attacking and increase its revenue. However, at Nash equilibrium, both attacks earn less than they would have if neither attacked. With multiple pools of equal size, a similar situation arises with asymmetric equilibrium.

Pontiveros et al. \cite{transSele} showed that the size and fee of the transaction have a higher importance in detecting mining pool strategies. Santos et al. \cite{efficentTrans} proposed a faster size-density table-based method that performs better in terms of the number of transactions processed and the total capital income. This approach is to remove sorting-based algorithms at block generation events. However, this method has not been compared with transaction selection strategies adopted by either public or private blockchains. Rizun \cite{blockSize} formalized the intuitive idea that the matching of supply with demand should determine equilibrium transaction fees. Fiz \cite{confirmationDelay} modeled the transaction selection problem as a classification problem and proved that the essential features of the transactions when selecting them are their size and fee values.

\section{Queuing Model Based Simulator}\label{sec-model}
In this section, a new queueing model for estimating transaction waiting time is proposed, based on which a simulator has been developed. The validity of the proposed model is checked with measurement data. 

By {\em transaction waiting time}, we mean the delay between when the transaction is received by the system and when the transaction is included in a block. Note that, there is additional delay till transaction confirmation, which is the delay for the system to achieve consensus and approve the addition of the block to the chain. Since this additional delay is not affected by the miners' transaction selection strategies, it will not be included in the model or later discussion if not explicitly stated.   

\subsection{Model Description}
\begin{figure}[ht!]
  \includegraphics[width=1\linewidth, height=0.18\linewidth]{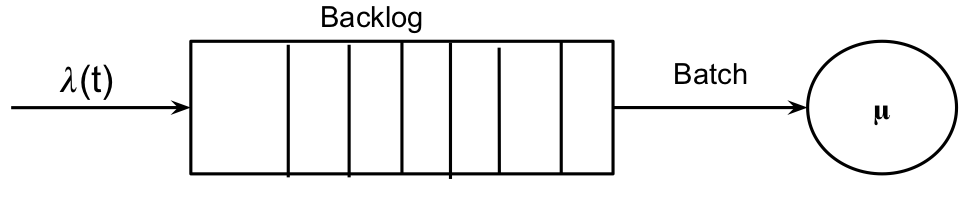}
  \caption{ M(t)/M$^N$/1 model}
  \label{model}
\end{figure}
The users generate transactions for processing, and the blockchain engine provides a secured, autonomous, and privacy-preserving platform.  The number of users that integrate the service increases exponentially, leading to the case in which the backlog gets filled with a large number of transactions waiting for the process.  Fig.~\ref{model} illustrates the Bitcoin workflow. In this case, taking into consideration the behaviour of a typical miner, we use a queueing model to represent the system. 
The users' newly generated transactions arrive at the system with an intensity of $\lambda (t)$, and the miners generate blocks with an intensity of $\mu$.  

\paragraph{Arrival process}
More specifically, the transaction arrivals follow an inhomogeneous Poisson process with an intensity of $\lambda (t)$ as having been observed in a measurement study \cite{TransBitcoin}. 

To generate the inhomogeneous arrival intensity from the homogeneous Poisson process, we can use the Lewis and Shedler thinning methodology \cite{lewisAndshelder},\cite{lewis}, as illustrated in Algorithm \ref{Algo}, where the $\lambda$ constrains the next arrival intensity. The $\lambda(t)$ is bounded by $\frac{\lambda(S_w)}{\lambda}$, where the $S_w$ is the next exponential inter-arrival time and $\lambda$ is the upper bound of $\lambda (t)$. Based on the current state of Bitcoin processing capacity, the value of $\lambda$ is set to 7.2 \cite{Btc}.  

\begin{algorithm}
\caption{Inhomogeneous Poisson Process}
\begin{algorithmic}[1]
\Procedure{Inhomogeneous}{$\lambda(t),T$}       
    \State Initialisation: $n=m=0, t_0=s_0=0$ 
    \State Condition: $\lambda(t) \leq \lambda,  \forall_ {t \leq T}$
    \While{$s_m \leq T$}  
        \State x $\sim$ U(0, 1) 
        \State $y=-\frac{ln(x)}{\lambda}$ 
        \State $s_{m+1}=s_m + y$ 
        \State $D \sim U(0, 1)$ 
        \If{$D \leq \frac{\lambda(s_{m+1})}{\lambda}$} 
        \State $t_{n+1}=S_{m+1}$ 
        \State $n=n+1$
        \EndIf
        \State \textbf{return} [$t_n$]
    \EndWhile  
\EndProcedure
\end{algorithmic}
\label{Algo}
\end{algorithm}

\paragraph{Arrival attributes} The new arrival transactions contain important features like fee and size that play a role in ranking order and filling up the block.  For instance, Bitcoin orders the new arrivals according to the fee per byte ratio. The weak dependency between the transaction fee and size impacts the number of transactions added to the block. In this work, we also introduce this dependency in the model. 

\paragraph{Service process}
The transactions are waiting at the backlog to be picked up and included in a block. Block generation is an independent and identically distributed random event requiring the miner to perform some mathematical puzzles, as Bitcoin's case.   The block-generation times follow exponential batch processing with a rate of $\mu$.  The block holds $N$ number of transactions, in which the size of the block ($\beta$) can only have as many numbers of transactions possible and available at the backlog. 
\paragraph{Block size}
A valid block holds $N$ number of transactions, and the maximum size of the block size ($\beta$) is fixed.  The pushing block size to the maximum limit also brings the propagation delay, which may trigger a fork in the distributed system.  However, it is crucial to see how the $\beta$ affects the transactions' average waiting time.  To see this effect, we compare block size from legacy size, which is 1 MB to 8 MB. 

\paragraph{Transaction selection / scheduling strategies}
To explore how much low-fee transactions may suffer from the strategy used by a miner in selecting / scheduling backlogged transactions in forming a block, three strategies are considered.  
One strategy we are considering is the fee per byte ordering at the backlog, which is the default strategy used by Bitcoin. 
In addition, the fee-careless first in first out 
(FIFO) strategy, and the strategy of prioritizing higher fees are also considered. 

 \begin{figure}[t!]
\centering
  \includegraphics[width=\linewidth, height=0.75\linewidth]{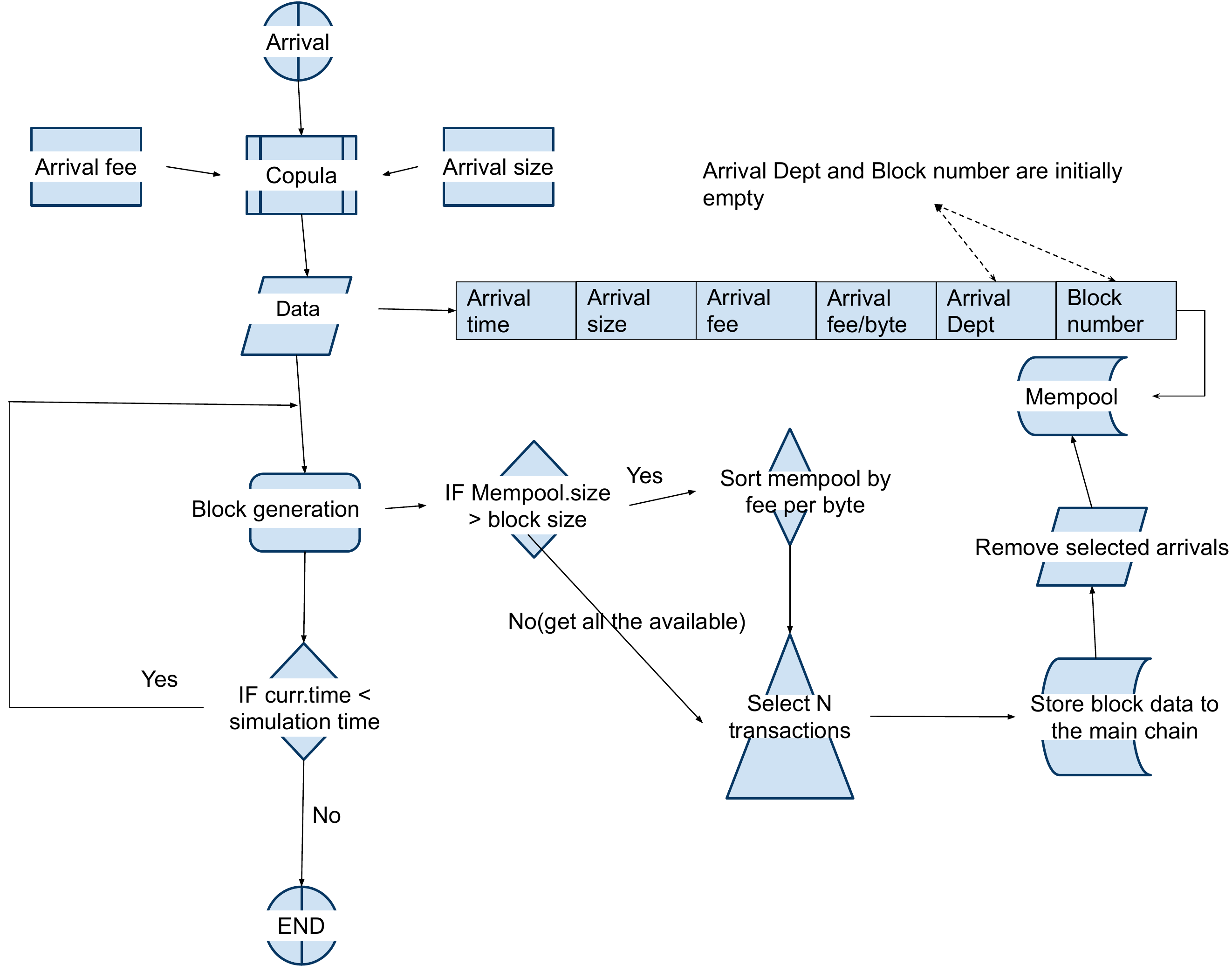}
  \caption{Flowchart showing the workflow of the model-based simulator} 
  \label{flowchart}
\end{figure}

\subsection{Simulator Workflow} \label{sec-sim}
This sub-section covers the workflow of the simulator. It captures the workflow of a full Bitcoin node that participates in the verification and validation of transactions. 
 
There have been some works on developing a simulator to study the evolving technology's performance, blockchain.  The currently available simulators focus on realizing node-to-node connectivity, propagation delay, and adding Merkle tree into the simulator, including \cite{Simb, simblock, blocksim, SIMBA, simarch}.  Since these simulators have no functionality to include the dependence between transactions fee and size, the change of the scheduling algorithms, and realizing inhomogeneous transactions arrival process, we developed a discrete event simulator/emulator by using Simpy \cite{simpy}.  

We can demonstrate the workflow of the simulator by using an example. Let a vector $[t, s, f, f/s, d, bN]$ represent a new arrival event at time $t$, in which the arrival has attributes of a fee ($f$), size ($s$), fee-per-byte ($f/b$), waiting time ($d$), and block number ($bN$). As it was discussed in previous sections, there is a weak dependence between arrival fee and size, this is realized by using Copulas \cite{copulas}.  Initially, the value of $d$ and $bN$ is zero.  Similarly, the other arrivals will be recorded at the mempool and waiting for a pick-up.  

When the block generation event happens, there are two ways of selection if there are enough arrivals stored for pick-up. Firstly, we can use the time of arrival of the transactions ($t$), which gives us FIFO. Secondly, we can consider the situation with the miners' knowledge having high incentives to increase the financial gain, prioritizing high fees (Fee-based). Thirdly, it uses the default method proposed by the Bitcoin community fee per byte ratio, $f/s$ order \cite{Nakamoto}. Then, the block's size and the associated transaction size determine the number of transactions included in a block.  If not, it picks-up the available arrivals and generates the block.   Fig. \ref{flowchart} illustrates the concept by considering fee per byte as the scheduling algorithm in the form of a flow chart at a high-level detail.  In this work, we consider fee-based, fee per byte, and FIFO.  The first two are used to show the impact of the miner incentives and FIFO to demonstrate the difference between financial gain or not.  

After the transactions are selected by one of the scheduling strategies mentioned earlier, the block containing the correspondingly selected transactions will be added to the chain. These transactions also get removed from the backlog.  The  chain grows in each block generation until the simulation window is finished.  Like the real Bitcoin node, this simulator keeps track of each transaction's arrival time, fee, size, block number, and waiting time.  These collected attributes are used to generate valid results and compare the result with currently available literature. 

The transaction arrival and block generation events change the state of the system. The arrival event increases the mempool in the number of arrivals and size-wise; on the other hand, the batch processing removes $N$ elements from the backlog.

\subsection{Model Validation} \label{sec-val}
This section presents results validating the model-based simulator with trace-driven simulation. 
It has been demonstrated \cite{TransBitcoin} that transaction fees and sizes follow a lognormal distribution with different mean and standard deviation while showing weak correlation, which is considered in the simulation. 

\begin{figure}[ht!]
\centering
  \includegraphics[width=0.8\linewidth, height=0.45\linewidth]{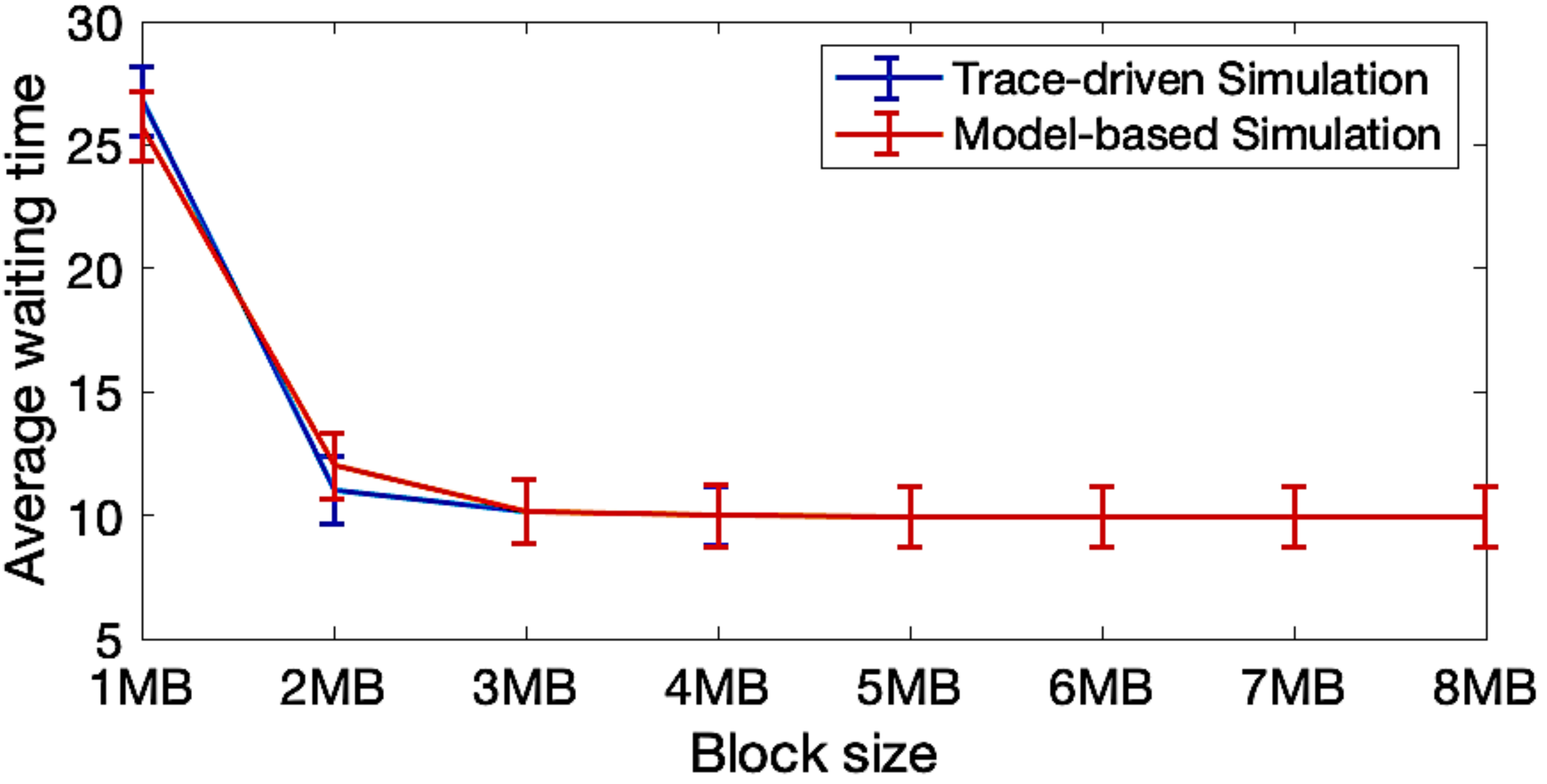}
  \caption{ Transactions average waiting time vs block size while fee per byte is the scheduling algorithm ($\lambda(t)\in [3.0, \dots, 3.3] $)} 
  \label{waiting}
\end{figure}

To validate the model, we performed a test and the results are reported in Fig. \ref{waiting}, where the model-based simulation results are compared with trace-driven simulation results.  As the figure illustrates, the model captures the results in a good fit.  The x-axis represents the block size. The y-axis indicates the transaction's average waiting time by considering the model and actual data from the bitcoin node. In this work, the average waiting time is time between a transaction generation and its addition to a valid block. As the default case in Bitcoin, a fee per byte is used to order the arrivals for pick up \cite{Nakamoto}. 

Table \ref{twominer} further illustrates a comparison between our proposed model with recent related works \cite{TransactionConfirmationBitcoin, trasactionConfirmation} and measurement results \cite{Banalysis}.  The row indicates the block size, and the column represents related models from the literature. The arrival transactions intensity is fixed with $\lambda = 3.0$. This table only considers block size from 1MB to 3MB. This is mainly because most paper commonly consider the block size from 1MB to 3MB, e.g., \cite{batchqueuing, DiscreteBlockchain}.  Our proposed model seems to fit with other related works' results. As a highlight, our model produces better matching result with the measurement  \cite{Banalysis}.

\begin{table}[ht!]
\centering
 \caption{Model comparison ($\lambda = 3.0$) } 
\begin{tabular}{|p{25mm}|p{10mm}|p{10mm}|p{12mm}|}
\hline
Models &  1MB   & 2MB & 3MB\\
\hline
$M/G^N/1$  \cite{TransactionConfirmationBitcoin, trasactionConfirmation} &  26   & 13.66 & 10.33\\
\hline
Bitcoin \cite{Banalysis} &  --  & 13.1  & --\\
 \hline
$M(t)/M^N/1$  & 25   & 13.01  & 10.14\\
 \hline
\end{tabular}
\label{twominer}
\end{table}

\section{Impact of Transaction Selection Strategy}\label{sec-res}
In this section, we investigate the impact of transaction selection / scheduling strategy used by a miner on transaction waiting time. The investigation is based on the simulator introduced in the previous sections. First, the validity of the law of conservation regarding scheduling algorithms in bringing the same average time is illustrated. Then, our simulation considers two cases that have been mentioned in Section \ref{sec-model}, {\bf (i)} the default method proposed by Bitcoin, which is the fee per byte, and {\bf (ii)} considering the particular case demonstrating the financial interest of the miner is only the fee.

\subsection{Conservation of Average Waiting Time}
\begin{figure}[b!]
\centering
  \includegraphics[width=0.8\linewidth, height=0.45\linewidth]{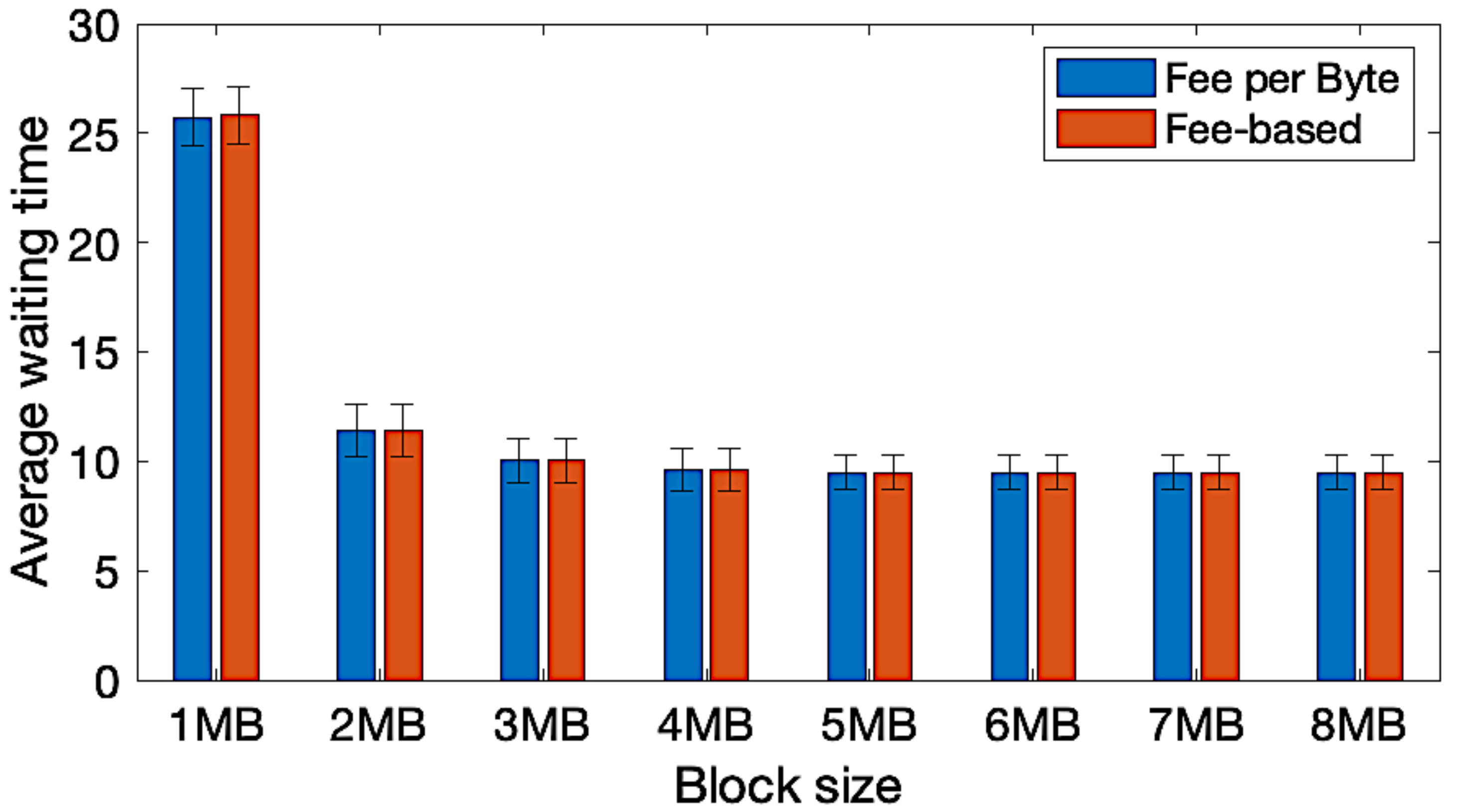}
  \caption{ Scheduling algorithms comparison ($\lambda(t)\in [3.0, \dots, 3.3] $)} 
  \label{sceduling}
\end{figure}

\begin{table}[b!]
\centering
 \caption{Filling rate comparison } 
\begin{tabular}{|p{14mm}|p{12mm}|p{13mm}|p{13mm}|p{13mm}|}
\hline
Strategies &  1MB($\mu,\sigma$)   & 2MB($\mu,\sigma$)  & 6MB($\mu,\sigma$) & 8MB($\mu,\sigma$)\\
\hline
Fee-based &  (0.86,0.02)   & (0.433,0.01) &(0.144,0.003)& (0.114,0.002)\\
\hline
Fee per byte  &  (0.85,0.021)  & (0.431,0.012)  &(0.143,0.003)& (0.111,0.002)\\
 \hline

\end{tabular}
\label{fillingrate}
\end{table}

Fig. \ref{sceduling} reports the average waiting time transactions seen while using fee per byte and fee-based.  The x-axis represents the block size in MB, the y-axis indicates the average waiting time, and the legend classifies the type of strategy used.  The plot illustrates that choosing any strategy while the arrival intensity is within the range of 3.0 to 3.3 may not affect the average waiting time.   However, this behavior can only apply when the number of arrivals waiting for pick up is smaller than the block can hold.  Table \ref{fillingrate} presents the filling rate of the block in terms of the mean and standard deviation.  The row represents the block size, and the column reflects the strategy applied.  As we can see from the table, the filling rate of the block in all the cases is lower than one, which means most of the time, the block is not pushed to maximum size. 

\subsection{Case-I (Fee per byte)}

Miners are the backbone of Bitcoin, participating in adding, validating, and forwarding new updates to the neighbors. Mainly, what a miner involves is solving the mathematical puzzle through high computation effort.  When the miner finds the nonce, it collects transactions from the backlog, ordering in fee per byte \cite{Nakamoto}.  In such cases, a transaction with a higher fee per byte ratio is picked up earlier than the low fee per byte. It is natural for the miners to choose transactions with a higher fee per byte since it increases the financial gain. However, this may affect the average waiting time for a low fee per byte transaction. It was demonstrated that the transaction fee fluctuates \cite{Btc}, transactions with a smaller fee observe a longer average waiting time \cite{batchqueuing}. There is a gap in the literature to illustrates how much a minor transaction has to wait. 

\begin{figure}[b!]
\centering
  \includegraphics[width=0.8\linewidth, height=0.45\linewidth]{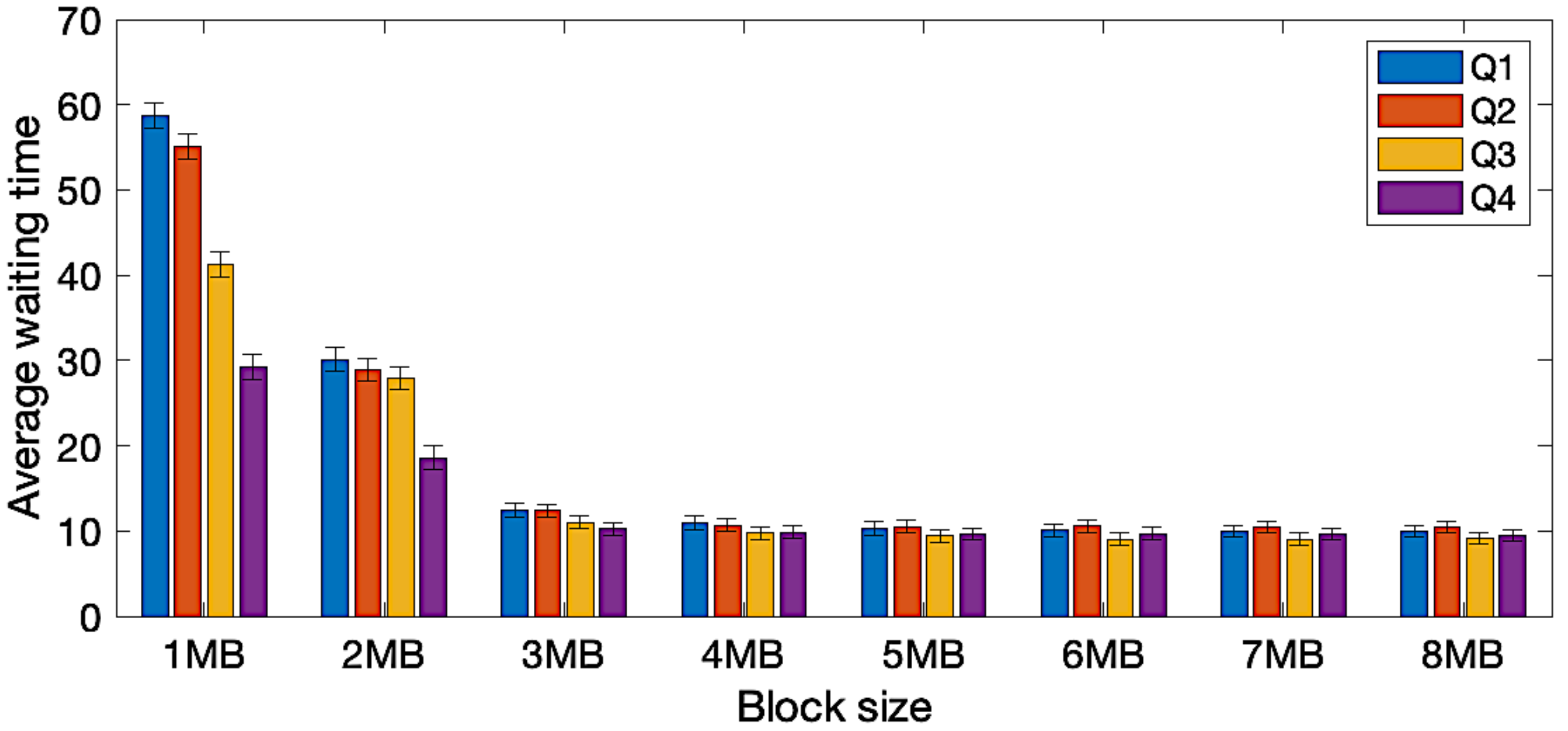}
  \caption{ Fee per byte ($\lambda(t)\in [7.0, \dots, 7.3] $)} 
  \label{CompFeeper}
\end{figure}

Fig. \ref{CompFeeper} illustrates the average waiting time seen by the transactions relative to the block size increase.  The x-axis represents the block size ranging from 1MB to 8 MB, and the y-axis shows the average waiting time.  It demonstrates the relationship between block size and average waiting time for Q1 (25\%), Q2 (50\%), Q3(75\%), and greater than Q3 ($>$Q3) for a fee per byte. As the figure shows, transactions with a low fee per byte ratio observe a higher waiting time. This is highly observable within the block size ranging from 1MB - 3MB. However, after 5MB, the effect of the financial incentive becomes smaller. This can also come because the mempool has fewer waiting transactions relative to the smaller block size, which forces the miner to pick up what is in the mempool. 

 \begin{figure}[b!]
\centering
  \includegraphics[width=0.9\linewidth, height=0.455\linewidth]{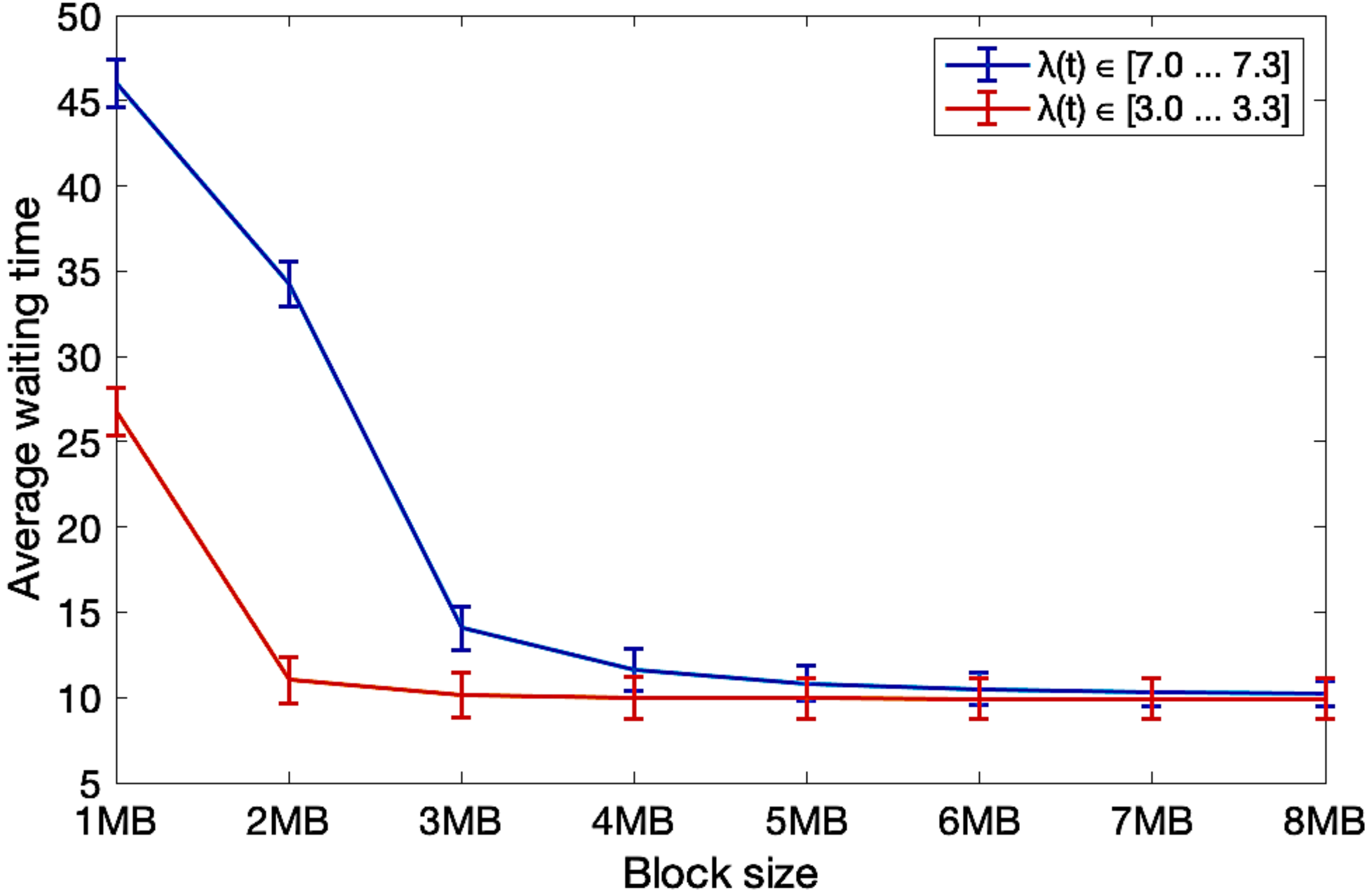}
  \caption{Arrival intensity  vs block size} 
  \label{high}
\end{figure}

Fig. \ref{high} reports the average waiting time a transaction sees while the block size is pushed to maximum and the arrival intensity are within range of 7.0 to 7.3.  As it is shown, when the block size increases, the average waiting time decreases.  Similarly, this trend is also visible when the intensity is within $\lambda(t)\in [3.0, \dots, 3.3]$. The reduction of the average waiting time after 6MB is smaller enough to be considered equal.

\begin{figure}[t!]
    \subfigure[Q1 ]
    {
    \includegraphics[width=0.45\linewidth,height=0.45\linewidth]{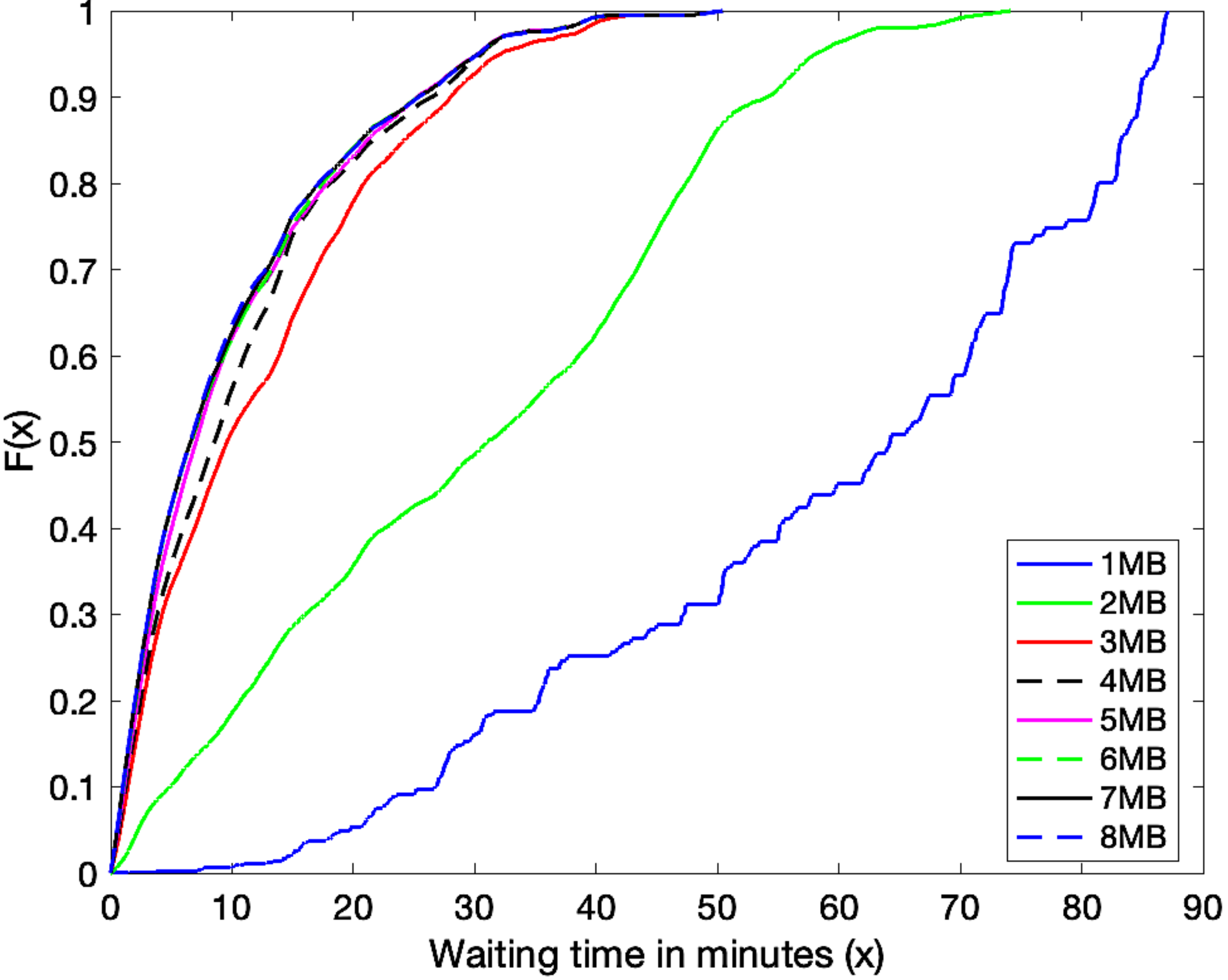}
   \label{Q11}
    }
    \subfigure[Q2]
    {
    \includegraphics[width=0.45\linewidth, height=0.45\linewidth]{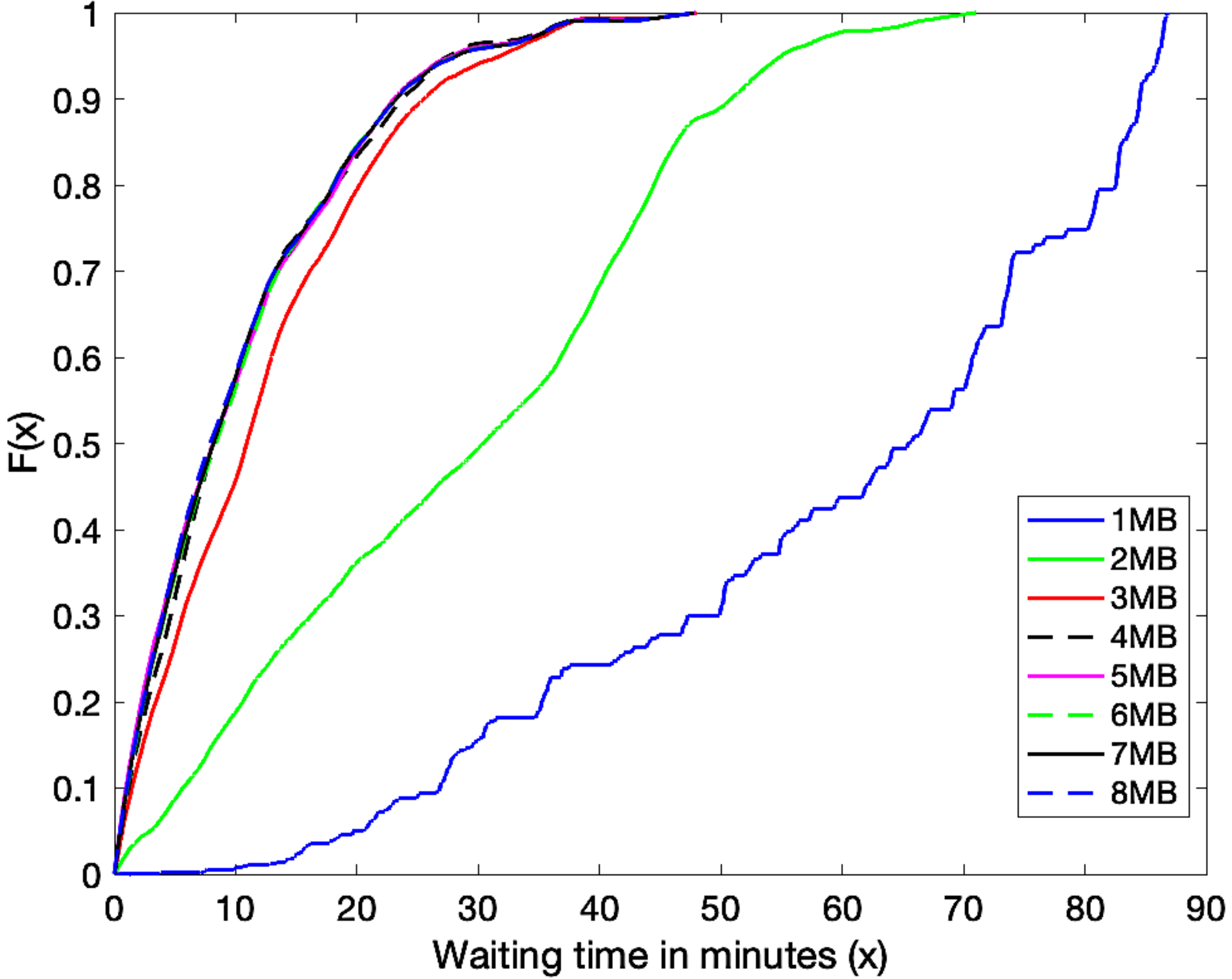}
   \label{Q22}
    }
    \subfigure[Q3]
    {
    \includegraphics[width=0.45\linewidth, height=0.45\linewidth]{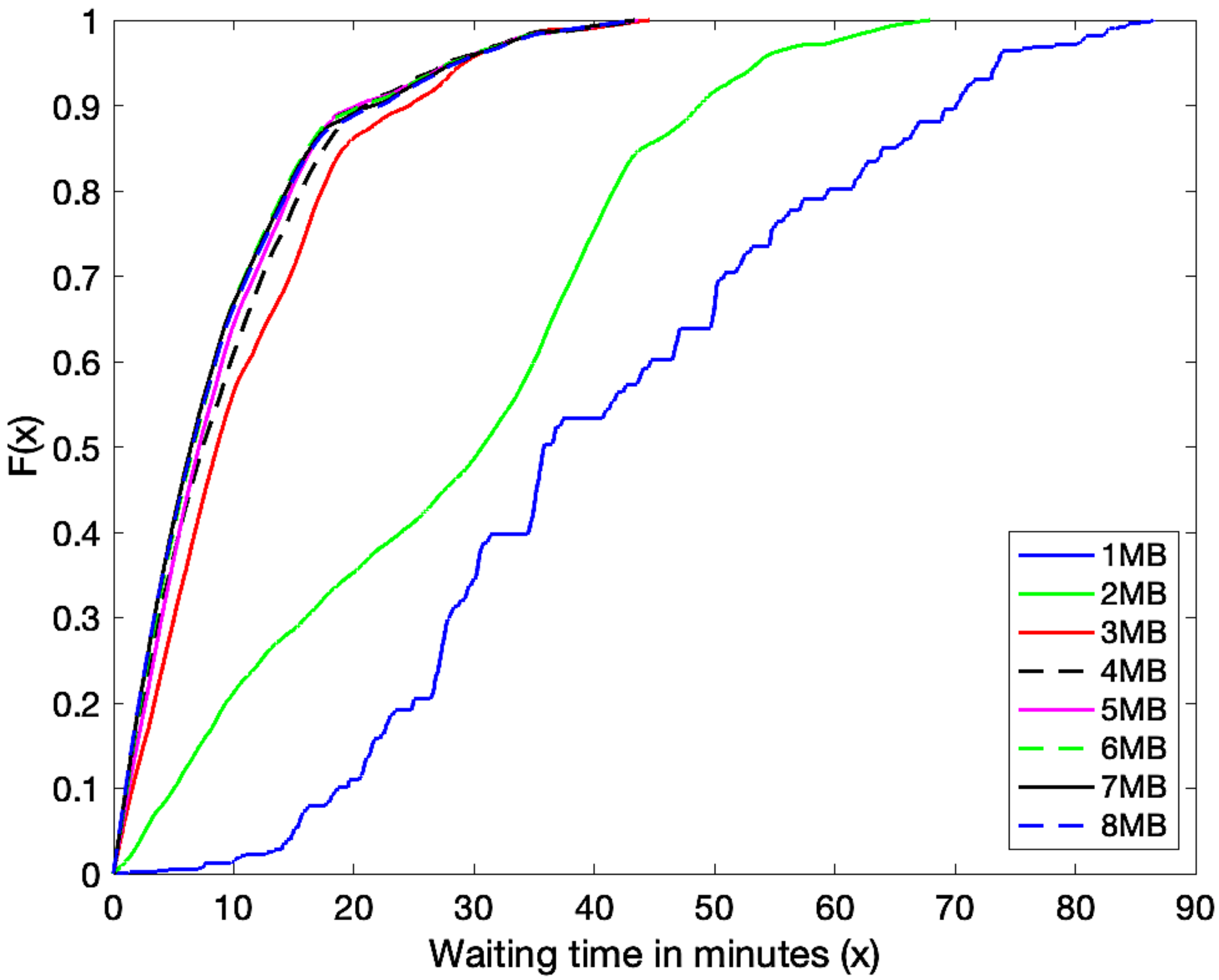}
   \label{Q33}
    }
     \subfigure[Q4]
    {
    \includegraphics[width=0.45\linewidth, height=0.45\linewidth]{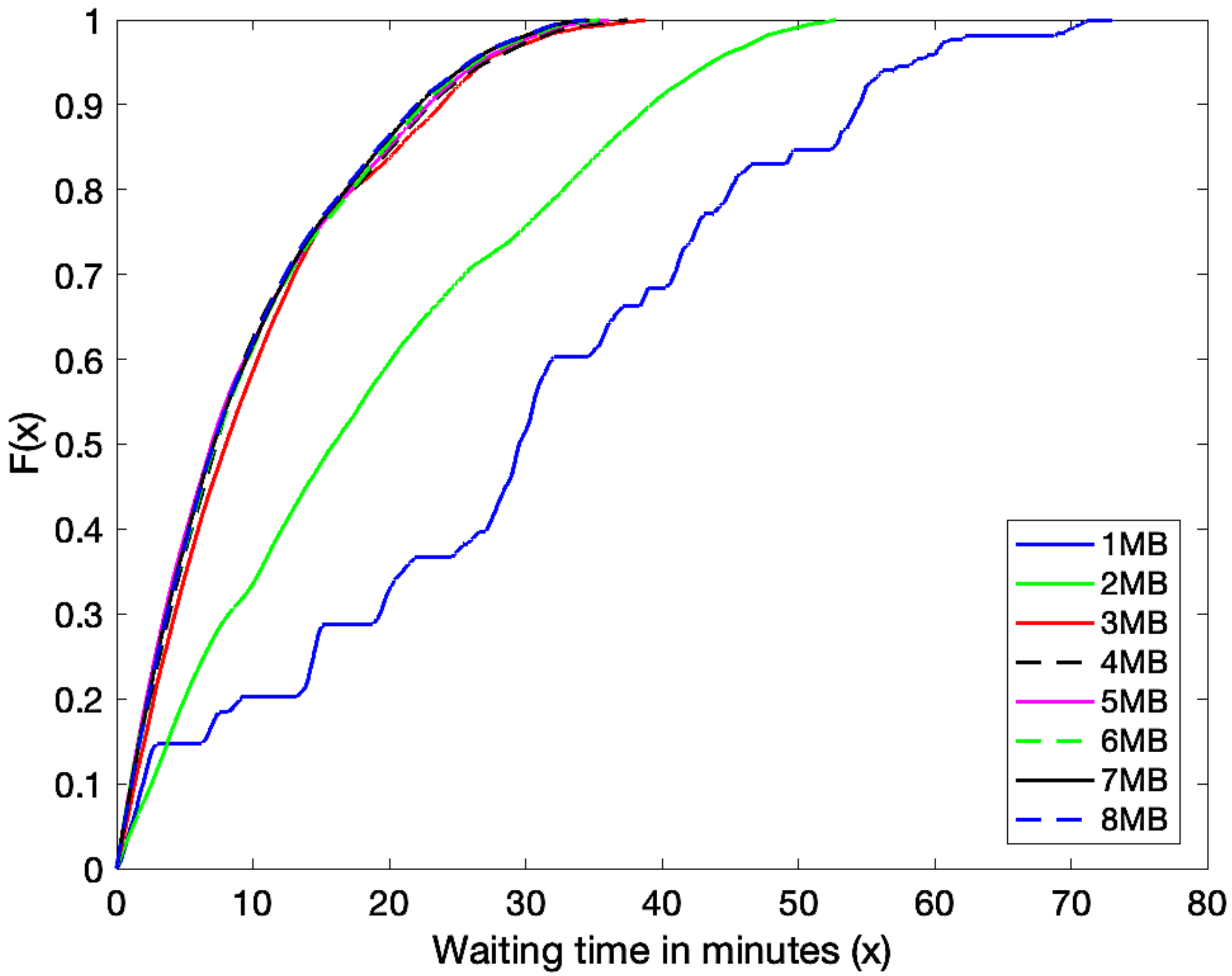}
   \label{Q44}
    }
  
  \caption{Transactions average waiting time vs. block size, where  $\lambda(t)\in [7.0, \dots, 7.3] $, fee per byte scheduling}  
  \label{Feeperbyte}
\end{figure}

Fig. \ref{Feeperbyte} reports the sample result showing the transaction waiting time in terms of low to a high fee per byte.  The x-axis represents the average waiting time in minutes. The y-axis is the empirical CDF. The legend in the plot classifies the transactions based on the block size. As we can see from the plot (\ref{Q11}) low fee per byte transactions, for 1MB block size, 80\% transactions see waiting time less than 70 minutes, and for 2MB, these transactions observe less than 45 minutes.

Similarly, for 3MB, smaller transactions 80\% of the time see less than 15 minute waiting time.  This behavior is repeated for a medium fee per byte (\ref{Q22}) size.  However, a high fee and very high fee per byte ratio transactions tend to see shorter waiting times.  For instance, in the case of very high fee per byte transactions, most transactions (80\%) see waiting time shorter than half an hour. Even increasing the block size has a more negligible effect on exhibiting the low-fee transaction suffering from longer waiting time.

\subsection{Case-II (Fee-based)}

We assume the miner prioritizes the financial motives over the default consideration of fee per byte \cite{minersfee}. A miner can sort the transactions in descending order of fee per byte, from the most profitable one to the least one \cite{feetrans}. By doing such ordering, it is easier to pick up a transaction that brings a higher profit.

\begin{figure}[t!]
\centering
  \includegraphics[width=0.8\linewidth, height=0.45\linewidth]{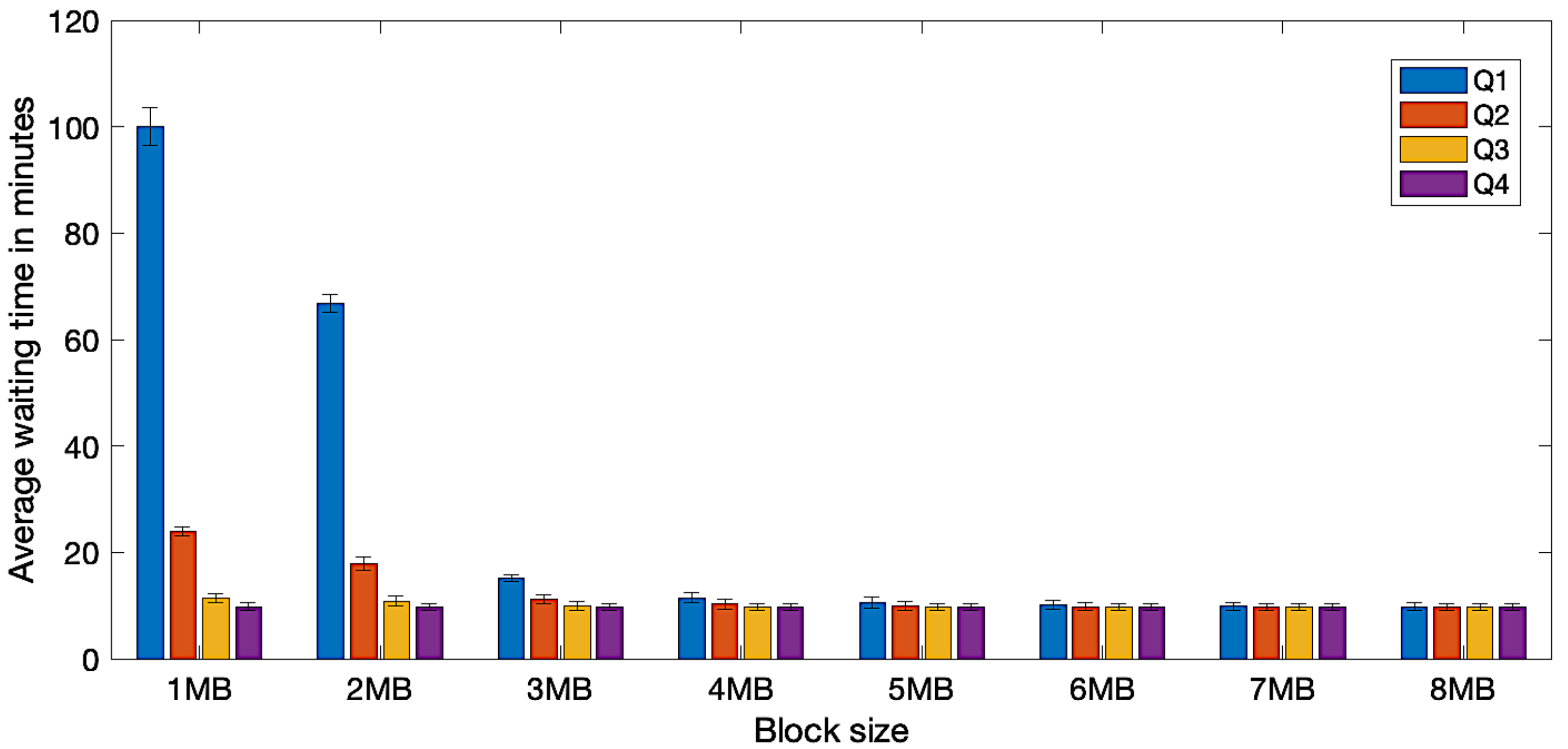}
  \caption{ Fee based $\lambda(t)\in [7.0, \dots, 7.3] $ } 
  \label{Compfee}
\end{figure}

Fig. \ref{Compfee} reports that miner financial interest is affecting the waiting time. For the block size from 1MB to 3MB, the impact of miners' incentives to select the top-fee transactions is more visible.  Transactions with smaller fees (Q1) wait for 30 minutes more than Q2.  However, starting the block size greater than 3MB, the average waiting time between smaller and higher becomes similar. 

\begin{figure}[t!]
\centering
  \includegraphics[width=0.9\linewidth, height=0.45\linewidth]{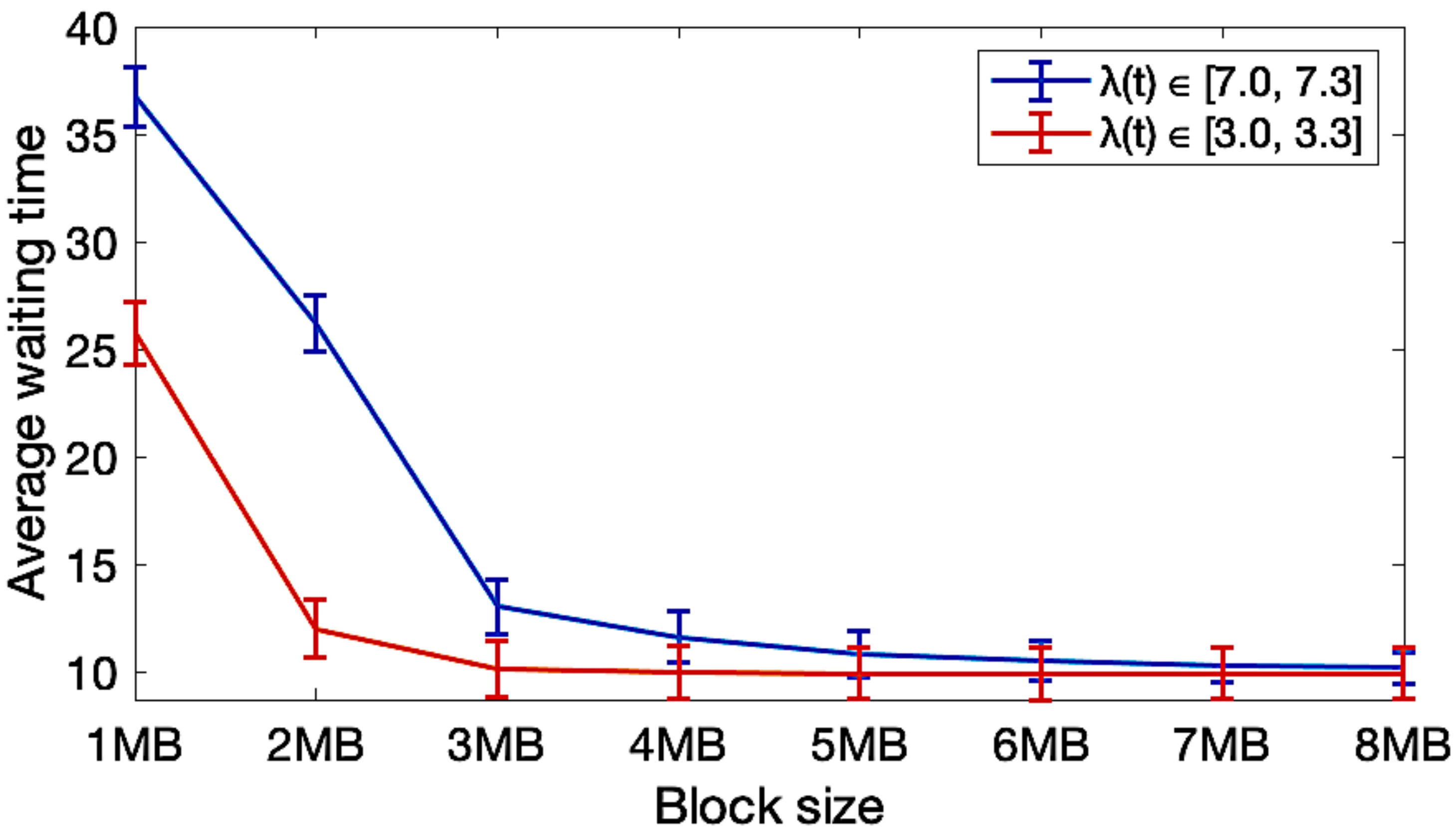}
  \caption{Arrival intensity  vs block size} 
  \label{highfee}
\end{figure}

Fig. \ref{highfee} reports the average waiting time seen by transactions when fee-based scheduling is used to pick up transactions from mempool.  The x-axis represents the block size, the y-axis indicates the average waiting time, and the legend classifies the two arrival intensity considered.  As we can see from the plot, when the arrival intensity is $\lambda(t)\in [7.0, \dots, 7.3] $, the average waiting time is seen by transactions when the block size is 1MB, or 2MB has smaller values than using fee per byte scheduling.  Starting 3MB, the average waiting time seen by using fee per byte or fee-based looks similar. 

\begin{figure}[th!]
    \subfigure[Q1 ]
    {
    \includegraphics[width=0.45\linewidth,height=0.45\linewidth]{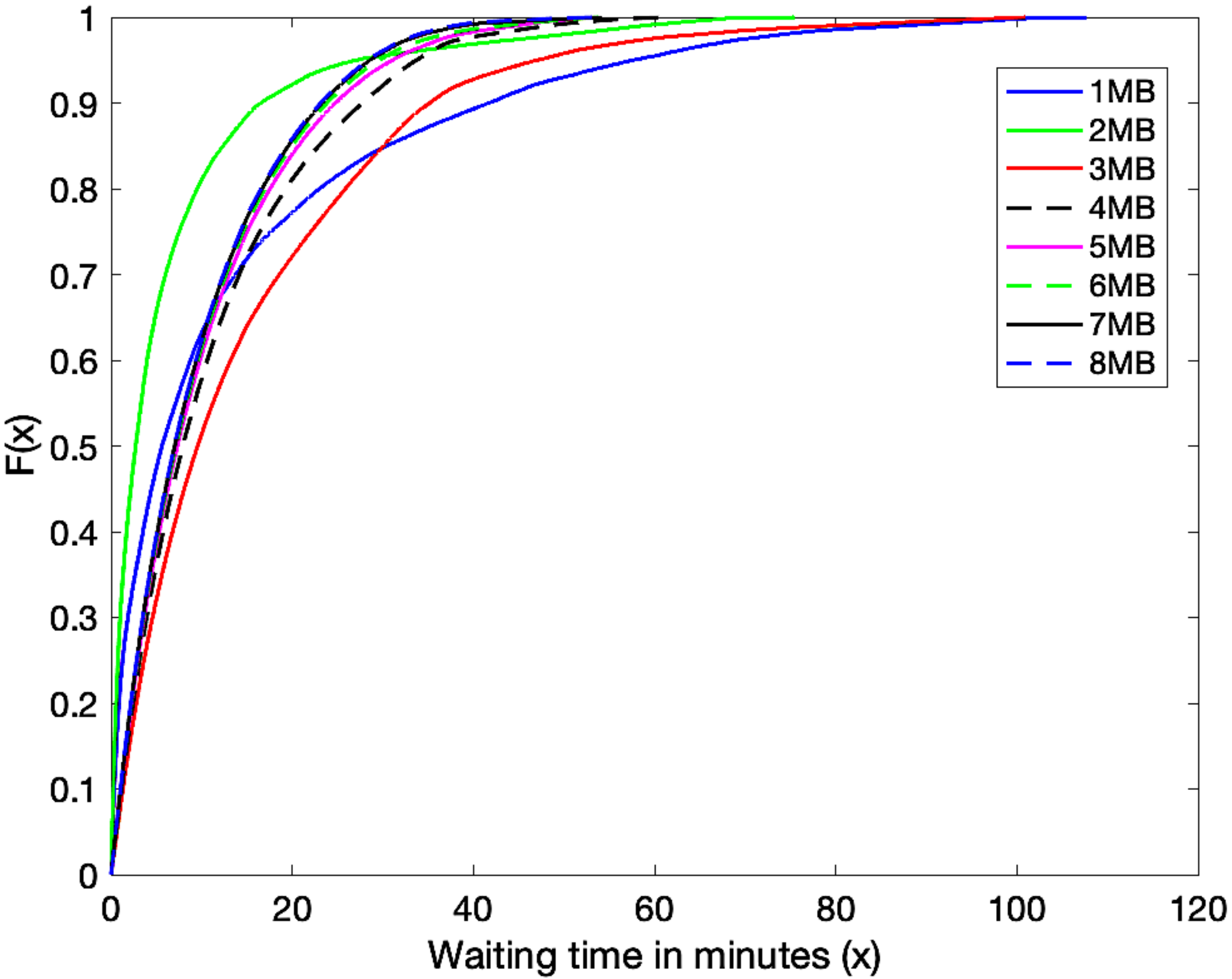}
   \label{Q66}
    }
    \subfigure[Q2 ]
    {
    \includegraphics[width=0.45\linewidth, height=0.45\linewidth]{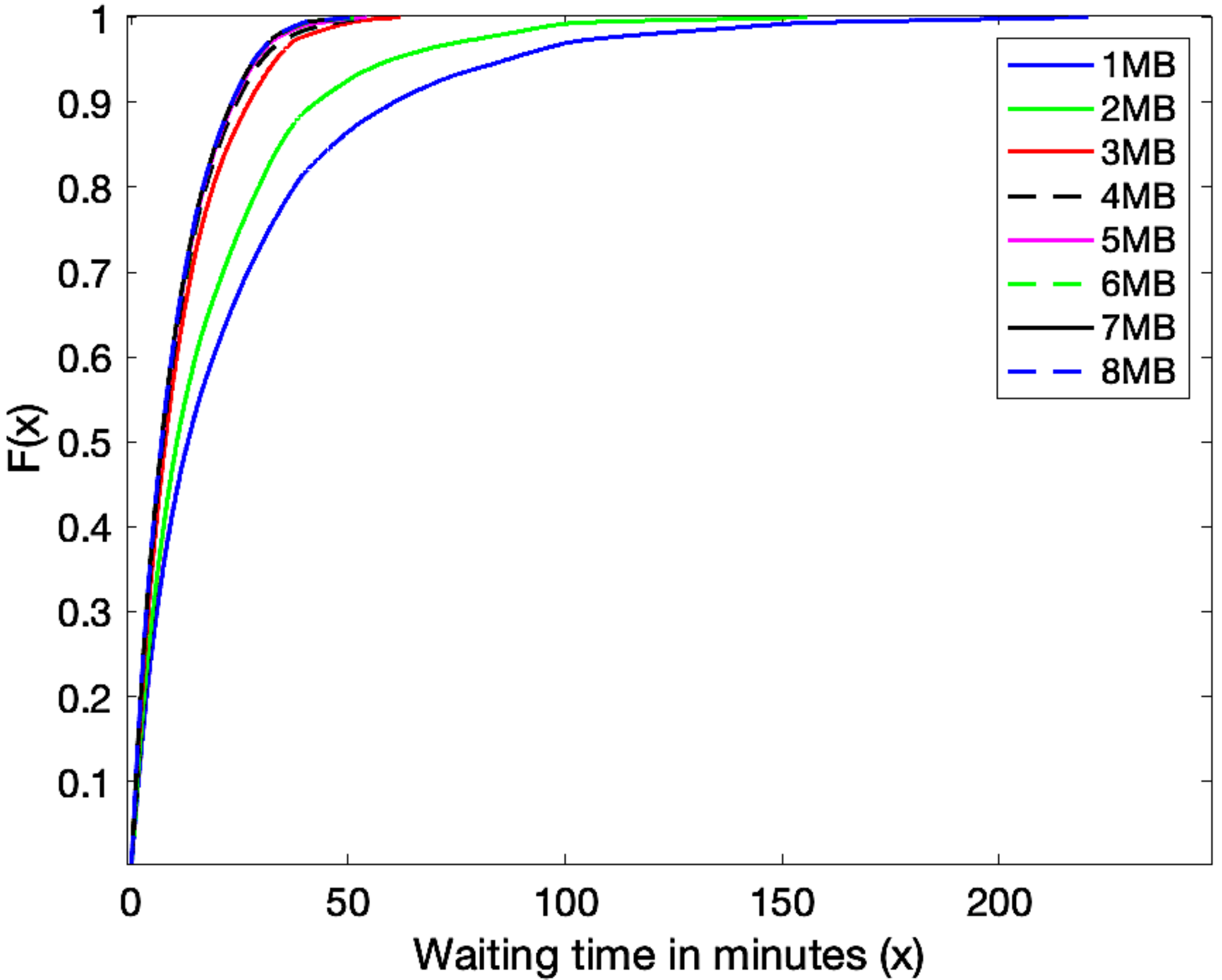}
   \label{Q77}
    }
    \subfigure[Q3]
    {
    \includegraphics[width=0.45\linewidth, height=0.45\linewidth]{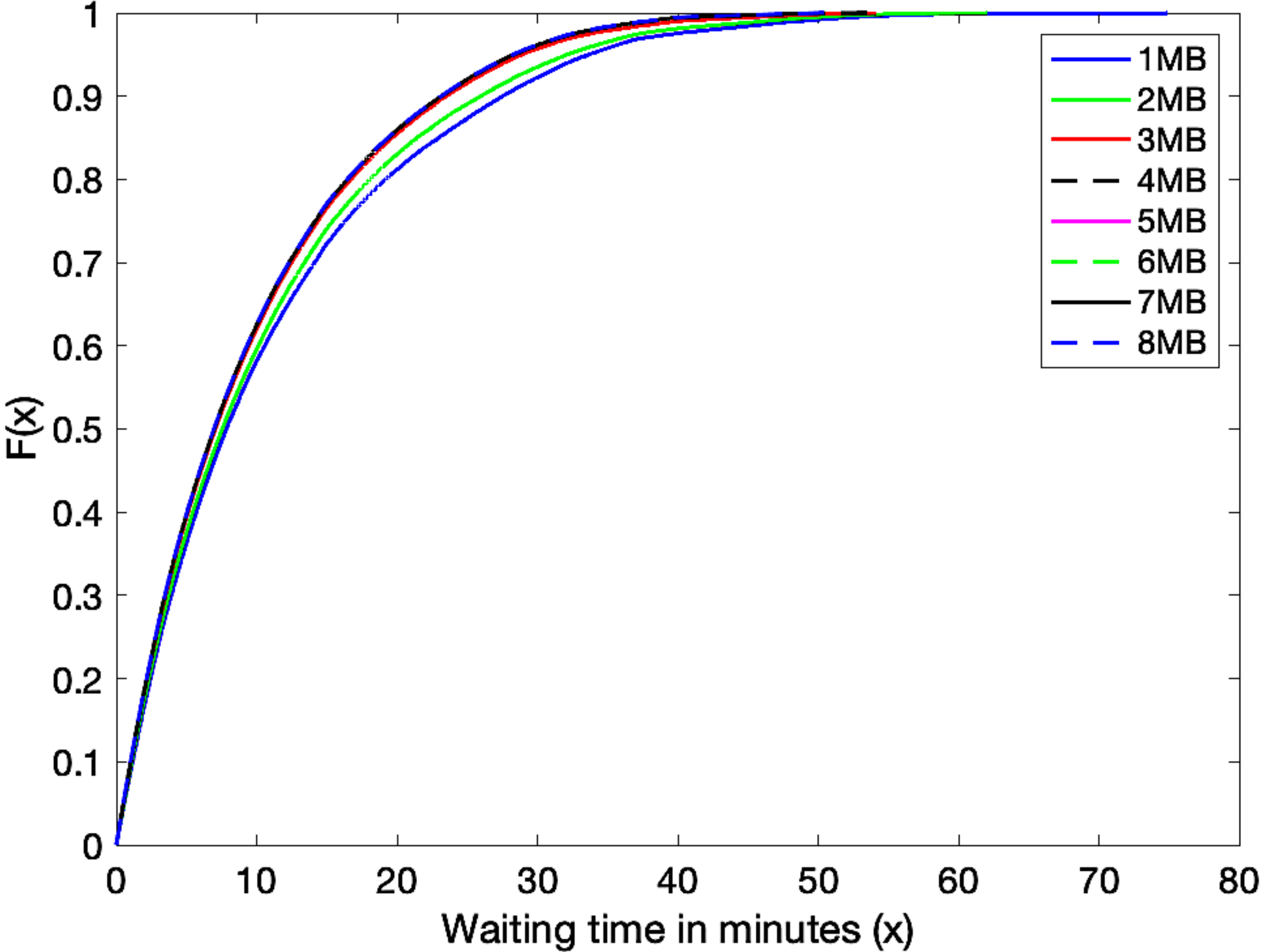}
   \label{Q88}
    }
     \subfigure[Q4]
    {
    \includegraphics[width=0.45\linewidth, height=0.45\linewidth]{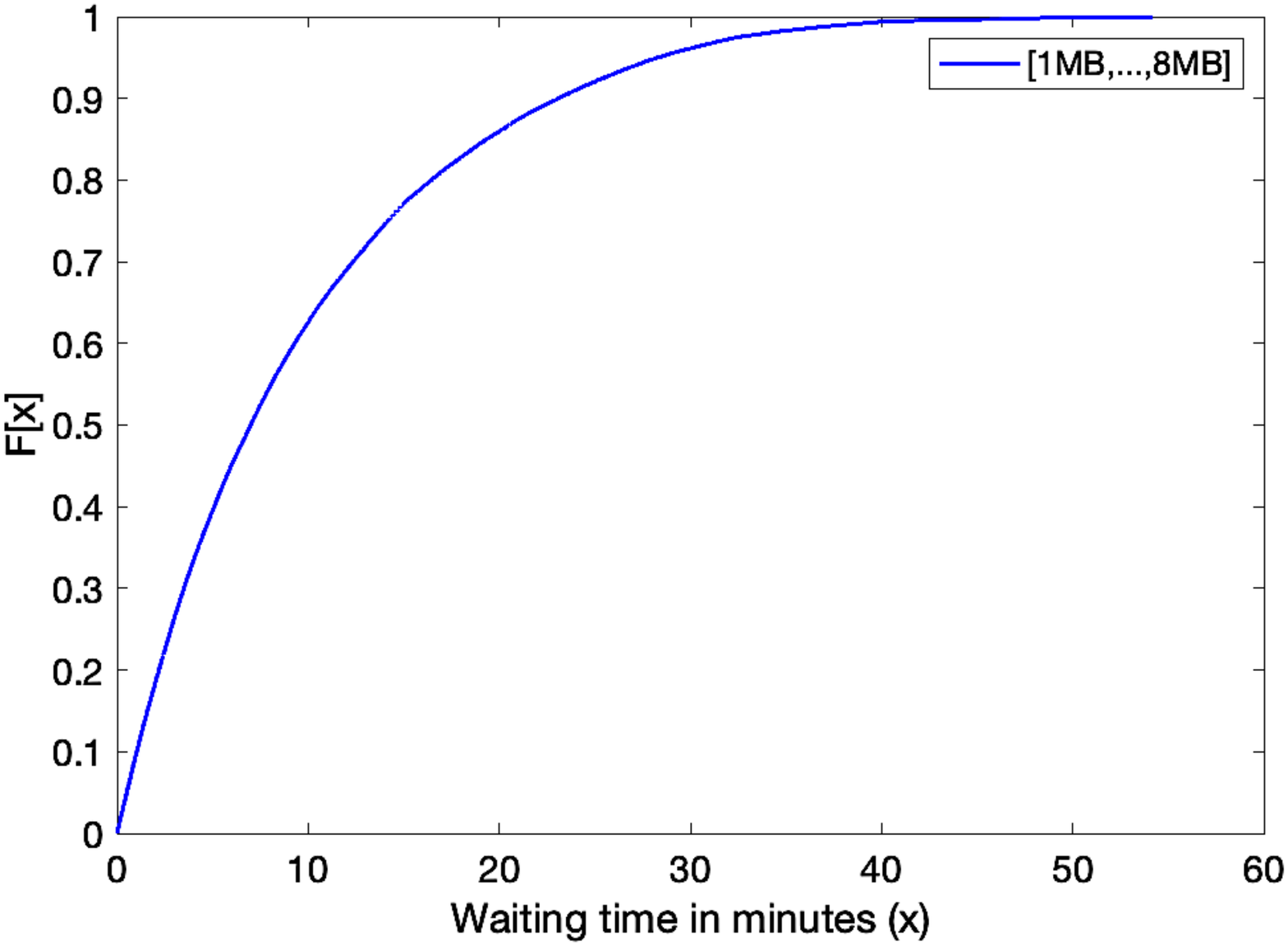}
   \label{Q99}
    }
  \caption{Transactions average waiting time vs. block size, where  $\lambda(t)\in [7.0, \dots, 7.3] $, fee based scheduling}  
  \label{feeBased}
\end{figure}

Fig. \ref{feeBased} shows how only choosing transactions with a higher fee affects the waiting time.  Fig. \ref{Q66} and \ref{Q77} plots values for 1MB, and 2MB are scale down by 10 for better visibility. Similar to the case where the fee per byte is used as the scheduling algorithm, higher fee transactions also show similar trends but different waiting times.  Since only choosing transactions with a higher fee does not consider the possibility of size limitations,  it seems fee per byte is a better option in giving fairer chances for transactions. However, this may change if the backlog is always full and there are more transactions to choose. In that case, selecting transactions based on fees may bring better gain but make low-fee transactions suffer long-time wait.

Unlike the fee per byte case, when transactions have a very high fee, they see the same average waiting time, implying the block size has a more negligible effect on the confirmation time.  90\% of the time, transactions see less than 25 minutes waiting time, and there are also less than 5\% of the transactions see more than 42 minutes average waiting time.

{\bf Remark:} The alert reader may have noticed that among the three strategies for transaction selection and block creation investigated in this study, we have left the FIFO out in the discussion above. This is simply because the fee-careless FIFO strategy does not show any significant difference between transactions of different fees or between transactions of different fees-per-byte: All transaction types have the same average waiting time.

\section{Reward Comparison} \label{sec-rew} 

\begin{figure}[hbt!]
\centering
  \includegraphics[width=0.8\linewidth, height=0.45\linewidth]{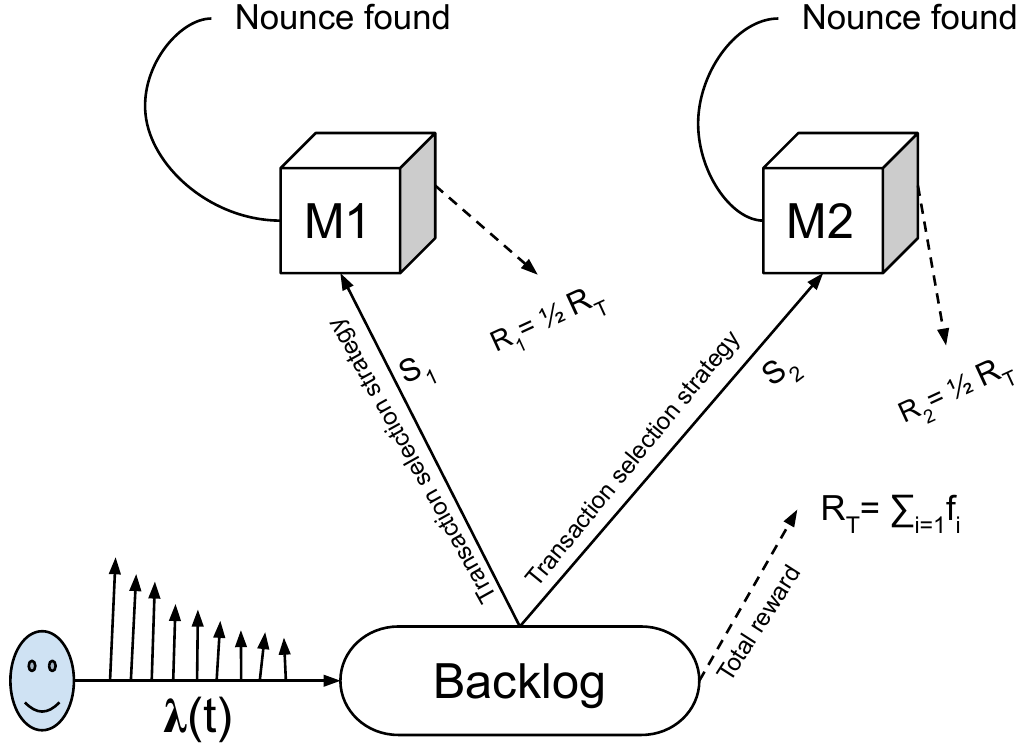}
  \caption{Miners} 
  \label{miners}
\end{figure}

In this section, we compare the strategy in terms of the reward a miner gets by adopting different strategies. We consider two miners (M1, M2) competing to generate a block with equal probability while using different strategies, as illustrated in Fig. \ref{miners}. These two miners share the same backlog. We consider each block generated by these miners valid and added to the main chain for simplicity. The total reward ($R_T=\sum_{i=1} f_i$) is the sum of all transactions' fee at the backlog. In a fair chance, each miner should get half of the reward ($\frac{1}{2} R_T$). We fix each miner's strategy and then compare each miners' total gain. We used the queue model-based simulator introduced in Section \ref{sec-model} to conduct the analysis.   

\subsection{Miner vs Miner}

The default strategy stated by the Bitcoin research community is fee per byte.  However, it is recommended but not enforced for the miners to follow this strategy. This enables the miners to choose a strategy that fits or increases the financial gain of the mining process, empowering the decision-making of miners to perform a   non-cooperative game.  In such a game, the Nash equilibrium states that a player can achieve the desired outcome by not deviating from their initial strategy \cite{Nash}. 

In the investigation of this subsection, we use the two-miner case to show if the Nash equilibrium exists. Tables \ref{stratfixed} and \ref{stratfixed2} illustrate the results from considering the two miners. These tables also show different block sizes while the arrival intensity is within the range of 3.0 to 3.3 ($\lambda(t)\in [3.0, \dots, 3.3] $),  to show the impact on the amount of gain by the miners. The values inside the table indicate the final gain of the miner while using one of the strategies.  For instance, row 2 and column 2 value (55.23, 55.23) means when both M1 and M2 use fee-based, they achieve the same financial gain. This gain is the sum of the transactions' fee picked by the miner utilizing this strategy.

 \begin{table}[t!]
\centering
 \caption{Strategy comparison (Block size=1MB)} 
\begin{tabular}{|p{15mm}|p{13mm}|p{15mm}|p{15mm}|}
\hline
Strategies &  Fee based   & Fee per byte & FIFO\\
\hline
Fee-based &  (55.23,55.23)   & (81.7,28.8)  & (67.6,42.85) \\
\hline
Fee per byte &(33.19,77.23)   & (55.45,55.46)   & (56.89,53.35) \\
 \hline
FIFO & (44.43,65.81)    & (57.67,52.55)   & (55.1,55.1) \\
 \hline
\end{tabular}
\label{stratfixed}
\end{table}

Table \ref{stratfixed} illustrates, there is a dominant strategy in this game for miner M1 and M2, i.e., use fee-based strategy. It is because the maximum payoff for row players in all columns occurs in the first row and first column. 
When M1 uses fee per byte, M2 maximum payoff occurs when it uses a fee-based strategy. Similarly, when M1 uses FIFO, M2 does best by changing into fee-based. However, M2's best strategy is not to change its current fee-based if M1 uses fee per byte or FIFO.

M1 and M2 have no incentive to change their strategy because fee-based is their dominant strategy. Since M1 uses fee-based in any case, M2's best response is not to change its fee-based strategy because it gets the maximum payoff. Given these facts, the cell gives us the maximum payoff for M2 in the first row. It is the first column that represents M2 not changing its fee-based strategy. Row 1 and column 1 hence shows a Nash equilibrium. 

 \begin{table}[t!]
\centering
 \caption{Strategy comparison (Block size=2MB)} 
\begin{tabular}{|p{15mm}|p{13mm}|p{15mm}|p{15mm}|}
\hline
Strategies &  Fee based   & Fee per byte & FIFO\\
\hline
Fee-based &  (55.24,55.24)   & (51.1,59.36)  & (52.4,58.05) \\
\hline
Fee per byte &(52.24,58.23)   & (55.7,55.7)   & (59.9,50.6) \\
 \hline
FIFO & (50.4,60.10)    & (57.70,52.78)   & (55.5,55.5) \\
 \hline
\end{tabular}
\label{stratfixed2}
\end{table}

 Table \ref{stratfixed2} demonstrates the impact in terms of final reward distributions. As the case for 1MB, M2 using a fee-based strategy is dominant in this case.  M2 achieves maximum payoff when the M1 uses the FIFO method. Similarly, M1 does better when M2 uses FIFO. Both M1 and M1 achieve the best when both use the same strategies. 

Since for miner M2, using a fee-based strategy is a dominant strategy, it makes  M2 have little incentive to change its strategy, which will leave M1 to also change to fee-based.  In this sense, column 1 and row 1 is a Nash equilibrium. 

\subsection{Miner vs Miners}
In this case, we considered five miners. Each miner has an equal probability of chance in generating a valid block and earning the reward. Four miners follow the same strategies while one miner chooses a different or the same strategy as the others. Same as the previous case, the arrival intensity is within range of 3.0 to 3.3 ($\lambda(t)\in [3.0, \dots, 3.3] $). Furthermore, the block size is fixed to 1MB or 2MB. In this section, M1 represents a miner with an independent incentive to change the strategy to increase the gain. However, M2 represents the other four miners following the same strategy while creating a block. Table \ref{strat2} and \ref{strat3} shows miners gain from adopting different block creation strategy. The values shown as (M1, M2) indicate the final gain of miner M1, and what each of  the other four miners earns. 

 \begin{table}[t!]
\centering
 \caption{Strategy comparison (Block size=1MB)} 
\begin{tabular}{|p{15mm}|p{13mm}|p{15mm}|p{15mm}|}
\hline
Strategies &  Fee based   & Fee per byte & FIFO\\
\hline
Fee-based &  (22.23,22.23)   & (35, 18.75)  & (32.2, 19.5) \\
\hline
Fee per byte &(10.13,25.1)   & (21.7,21.7)   & (23.2,21.75) \\
 \hline
FIFO & (13.13,24.2)    & (21.5,22.2)   & (21.6,21.6) \\
 \hline
\end{tabular}
\label{strat2}
\end{table}

Table \ref{strat2} and \ref{strat3} demonstrate that using a fee-based strategy increases the gain of single or grouped miners. When a miner uses this strategy, it achieves better gain than following another strategy. However, when all five miners use the same strategy, the gain is equally divided. This result shows that when all the miners follow the same strategy, the reward is equally divided. Otherwise, miners can adopt different strategies to increase financial gain. This implicitly encourages miners to adopt or change their strategies to achieve higher financial gain, regardless of transactions that give smaller gains may take longer to be processed than expected. 

\begin{table}[t!]
\centering
 \caption{Strategy comparison (Block size=2MB)} 
\begin{tabular}{|p{15mm}|p{13mm}|p{15mm}|p{15mm}|}
\hline
Strategies &  Fee based   & Fee per byte & FIFO\\
\hline
Fee-based &  (22.23,22.23)   & (25, 21.25)  & (22.2, 21.95) \\
\hline
Fee per byte &(20.9,22.27)   & (22,22)   & (24,21.5) \\
 \hline
FIFO & (22.13,21.96)    & (18.5,22.88)   & (22.1,22.1) \\
 \hline
\end{tabular}
\label{strat3}
\end{table}

\section{Discussion }\label{sec-dis}

There has been some research work proposing schemes and methods in increasing the throughput of Bitcoin \cite{scalab}.  These proposals focus on either increasing block size \cite{blocksize, blocksizeGame} or validating transactions outside of the main chain \cite{offchain, offchainTrust}.  From the Bitcoin design perspective, the average inter-block generation time is 10 minutes, making the previous blocks reach all the nodes in the network \cite{Nakamoto}. 
Around 18.5 million mined bitcoin are circulating on the network as of July 2020\footnote{https://www.blockchain.com/charts/total-bitcoins}. Since its inception in 2008, there has been a growing interest in studying Bitcoin.  Despite its popularity, slow transaction processing speed is one of the fundamental issues that make Bitcoin struggling to address. The ever-increasing issue of smaller fee transactions waiting a long time to be processed was started around April 2017 and is still not addressed.  In around April - August 2017, the throughput's reductions had been very steep. The Bitcoin community was forced to extend the block size by 1 MB to reduce the number of transactions waiting for confirmation.  In \cite{Btc}, we can also observe the increase of the throughput monotonically after soft-fork extensions. However, based on the number of applications that integrate the bitcoin service, there is no guarantee we will not face the same issues in the future. 
Increasing the block size may require more than the default average inter-generation time to propagate.  As reported in Fig. \ref{Feeperbyte} and \ref{feeBased}, increasing the block size may process more transactions per block, but the low-fee transactions still suffer from a long waiting time.  Increasing the block size increases the block propagation time and impacts the consistency of the ledger \cite{arthur}. In addition,  the backlog of transactions awaiting inclusion in future blocks will clog up the bitcoin network. The bitcoin nodes which form the collective backbone that relays transactions across the network, will be overloaded with data, and some transactions could be severely delayed or even rejected altogether. Similarly, shortening the block generation interval increases the fork rate in the system, which compromises the platform's security \cite{arthur}.  Hence, the main issue resides in the proper management of the backlog, which requires an independent investigation to improve the technology's quality of service. 

\section{Conclusion} \label{sec-con} 
In this paper, we analyzed the transaction waiting time for Bitcoin. Specifically, we modeled the transaction waiting time process as a single server with batch processing and different transaction selection strategies.  We considered that transaction priority is only dependent on the transaction fee and size. To study the transaction waiting time,  we developed a single node simulator/emulator that captures the workflow of Bitcoin. The proposed model shows that transactions with a minimal fee or fee per byte sufferers from a long waiting time even with the maximum block size.  

In addition, we performed analysis on the impact of a miner's transaction selection strategy on the final gain or loss.  The analysis shows that when miners use the same strategy, the average income between the miners is equally divided. However, when miners choose a different strategy, they can achieve different gain relative to the opponent strategy. Other than the transaction selection strategy, we also showed that the block size also impacts miners to choose which method to choose from, mainly because it decides the number of transactions. These results show that increasing block size alone may not bring optimal solutions. Performing an independent investigation on the backlog to introduce fairness in terms of waiting time to the minor transactions is needed.

\AtNextBibliography{\footnotesize}
{\footnotesize \printbibliography}

\end{document}